# Scrutinizing the double superconducting gaps and strong coupling pairing in (Li$_{1-x}$Fe$_x$OH)FeSe

Zengyi Du*, Xiong Yang*, Hai Lin*, Delong Fang, Guan Du, Jie Xing, Huan Yang†, Xiyu Zhu, and Hai-Hu Wen†

Center for Superconducting Physics and Materials, National Laboratory of Solid State Microstructures and Department of Physics, Collaborative Innovation Center for Advanced Microstructures, Nanjing University, Nanjing 210093, China

*These authors contributed equally to this work.

**In the iron based superconductors, one of the on-going frontier studies is about the pairing mechanism. The recent interest concerns the high temperature superconductivity and its intimate reason in the monolayer FeSe thin films[1-4]. The challenge here is how the double superconducting gaps[1] seen by the scanning tunnelling spectroscopy (STS) associate however to only one set of Fermi pockets seen by the angle resolved photoemission spectroscopy (ARPES)[3,4]. The recently discovered (Li$_{1-x}$Fe$_x$OH)FeSe phase[5-7] with $T_c \approx 40$ K provides a good platform to check the fundamental problems. Here we report the STS study on the (Li$_{1-x}$Fe$_x$OH)FeSe single crystals. The STS spectrum clearly indicates the presence of double anisotropic gaps with maximum magnitudes of $\Delta_1 \approx 14.3$ meV and $\Delta_2 \approx 8.6$ meV, and mimics that**

**of the monolayer FeSe thin film. Further analysis based on the quasiparticle interference (QPI) allows us to rule out the *d*-wave gap, and for the first time assign the larger (smaller) gap to the outer (inner) hybridized Fermi pockets associating with the $d_{xy}$ ($d_{xz}/d_{yz}$) orbitals, respectively. The huge value $2\Delta_1/k_BT_c$ = 8.7 discovered here undoubtedly proves the strong coupling mechanism in the present superconducting system.**

The pairing mechanism and the superconducting gap structure have remained as the core issues in the study on iron based superconductors. Depending on the atomic structures, the Fermi surface topology changes significantly in different systems, varying from hole pocket dominated, to both hole and electron pockets, and to only electron pockets in the FeSe based systems[1-7]. In many earlier discovered FeAs-based systems[8-10] with both electron and hole pockets, a natural paring picture was proposed[11,12], which concerns the pairing through the pair-scattering process between the electron and hole pockets leading to an attractive pairing potential and the pairing manner of $s^{\pm}$. While in the system with absence of the hole pockets, this weak coupling based picture is facing a great challenge. The question arising immediately is that whether we have a sign change of the gap function as expected from the repulsive interaction induced pairing[13], for example the nodeless *d*-wave, between two neighboring electron pockets. The recently discovered (Li$_{1-x}$Fe$_x$OH)FeSe phase[5-7] with $T_c \approx$ 40 K shows also only electron pockets around $\widetilde{M}$ points in folded Brillouin zone (BZ)[14-15], which is very similar to the case of monolayer FeSe thin film. Here we use the STS measurements and the QPI analysis to carefully scrutinize the superconducting gaps on the associated Fermi surfaces in (Li$_{1-x}$Fe$_x$OH)FeSe.

The (Li$_{1-x}$Fe$_x$OH)FeSe single crystals were made by following the hydrothermal ion-exchange method[5-7] based on starting crystals of K$_x$Fe$_{2-y}$Se$_2$. The (Li$_{1-x}$Fe$_x$OH)FeSe single crystal shows perfect Meissner shielding and sharp resistive transition. The details of growth and characterization can be found in the Supplementary Information (SI).

Scanning tunnelling microcopy/spectroscopy (STM/STS) measurements provide key information on the electronic structure and the gap symmetry of the novel superconductors[16-18]. Figure 1a shows a topographic image of the terminated surface of the cleaved (Li$_{1-x}$Fe$_x$OH)FeSe single crystal, and the atomically resolved square lattice is clearly seen. The lattice constants at the perpendicular directions are 3.61 Å and 3.63 Å, respectively, which are comparable to the lattice constant *a* of 3.79 Å from previous reports[5,7]. The schematic structure of (Li$_{1-x}$Fe$_x$OH)FeSe is shown in Fig. 1b. Since the cleaved surface is very stable and the lattice constant is very close to the expected value of Se-Se bond, together with the perfect STS measured (see below), we can reasonably assume that the terminated top layer is the Se-atom layer. Some bright spots with a dumbbell shape on the topography are observed and very similar to those from Cu or Co impurities in our previous works[19,20]. From the structure of the system, it is found that the impurity atom in the center of a dumbbell spot locates just at the position of Fe atoms in the beneath layer, this may be induced by the partial substitution of Fe atoms with Li, or some vacancies at the Fe sites in the Fe layer. The calculated density of the impurities varies from 0.5% to 1.5% of Li(Fe) sites, which is much lower than the ratio of about 20% of substitution of Li by Fe atoms in the Li(Fe) layers[5-7].

Fig. 1c shows a typical tunnelling spectrum measured at 1.5 K. Clearly there are two pairs of coherence peaks on the spectrum, indicating existence of two superconducting gaps. The larger gap is marked by $\Delta_1$ on its peak position of about ± 14.3 meV while the smaller one is marked by $\Delta_2$ at about ± 8.6 meV. The spectrum with two-gap feature is reminiscent of that reported on the monolayer FeSe thin film on $SrTiO_3$ substrate[1]. The recent high resolution ARPES measurements show only electron pockets as well as one superconducting gap in $(Li_{1-x}Fe_xOH)FeSe$[14-15]. It should be noted that the two-gap feature in the monolayer FeSe thin film was observed only in the STS measurements[1], not in the ARPES. This may be induced by the limited resolution of ARPES, or due to some selection rules in the emission process of the photo-electrons. More interestingly, the gaps found by the two ARPES measurements from two different groups on the similar samples are corresponding to the larger[14] and smaller gap[15] maximum of our STS measurements, respectively. The two-gap feature is a common one in our $(Li_{1-x}Fe_xOH)FeSe$ samples, which can get support from the statistical result of the peak positions on about 100 spectra collected in our experiment, which is shown in Fig. 1d. By fitting to Gaussian functions, we obtained the two gap values of $\Delta_1$ = 14.3 ± 1.3 meV and $\Delta_2$ = 8.6 ± 0.9 meV, and the corresponding ratios of $2\Delta_1/k_BT_c$ = 8.7±0.8 and $2\Delta_2/k_BT_c$ = 5.2±0.5. These values are much larger than the one $2\Delta/k_BT_c \sim 3.53$ predicted by the Bardeen-Cooper-Schrieffer (BCS) theory in the weak coupling regime. The huge value $2\Delta_1/k_BT_c$ = 8.7 indicates very strong coupling mechanism in the system. Worthy of noting is that similar ratio was observed in the cuprate superconductors[21,22].

In Fig. 2a, we show the temperature evolution of the spectra measured at the same position. The coherence peaks corresponding with both gaps are suppressed and mix together with increasing temperature, and finally vanish above $T_c$. Shown in Fig. 2b are the normalized spectra (divided by the one measured at 40 K). The two well resolved coherence peaks together with the fully gapped density of states (DOS) near zero-bias allow us to make reliable fitting using the Dynes model[23]. We firstly try to fit the STS by using two isotropic s-wave gaps, but it fails to catch up the main features of the experimental data as shown in Fig. S1 of SI. We then use the two anisotropic s-wave $\Delta_{1(2)}(\theta) = \Delta_{1(2)max}[1 - p_{1(2)}(1-\cos 4\theta)]$ to fit the experimental data. From Fig. 2b one can see that the theoretical model fits the experimental data very well. The fitting process is detailed in SI(II) and the fitting parameters are listed in SI-Table S1. The fourfold symmetric gap functions used to fit the data at 1.5 K are presented in Fig. 2c, and the anisotropy $p$ is fixed in all the fittings, i.e., 25 % for $\Delta_1(\theta)$ and 15% for $\Delta_2(\theta)$, while the weight of each gap changes at different temperatures. The resultant gap maximum versus temperature is shown in Fig. 2d, and the solid lines are the result of theoretical calculation within the BCS model by fixing $\Delta_{max}(0)$ and $T_c$ derived from our experiment. One can see that the temperature dependent superconducting gaps can be well described by the BCS theory. Our results suggest that this material have two nodeless superconducting gaps, each with a significant anisotropy. The two-gap feature is difficult to be understood if only one set of electron Fermi pockets around the $\widetilde{M}$ point exists as observed in the ARPES measurements.

On the surface of the crystal, we can see some dumbbell-shape impurities. The spectrum measured on the impurity site has no well-defined superconducting coherence peaks as shown in SI-Fig. S3. Instead, we can see the in-gap states with two asymmetric peaks which make the bottom near zero-bias "V"-shaped. The impurities which act as the scattering centers will produce the standing waves, and such quasiparticle interference (QPI) imaging can provide fruitful information on the scattering between the contours of Fermi surfaces with the scattering vector $\vec{q}$ at constant energy $E$ in $\vec{k}$-space[18]. The standing waves can be clearly seen in Fig. 3a measured around the larger energy gap $\Delta_1$. When we do the Fourier transformation (FT) to the QPI image, the resultant $Z(\vec{q}, E)$ image, or called as the FT-QPI image, shown in Fig. 3b helps us to investigate the band structure and the superconducting gap of the material. The ARPES[14,15] measurements show only one set of electron pockets near $\widetilde{M}$ points. However if we take a closer look at the center of the FT-QPI images as shown in Figs. 3f and 3g, we can find that it is unlikely there is only one ring in the center, but two rings, which means that a single set of isotropic ring-shaped Fermi surfaces would not be suitable for producing such pattern. Therefore we use the anisotropic electron Fermi surfaces (Fig. 3d) to simulate the QPI image, and the resultant pattern (Fig. 3e) is comparable with our experimental data. From our experimental data, there is almost no intensity at a $\vec{q}$-vector which connects the $\Gamma$ and $\widetilde{M}$ point in the FT-QPI image, this suggests no Fermi surface nor DOS near the $\Gamma$ point. As shown in Figs. 3b and 3c, when the bias voltage was increased from $\Delta_{2max}$ = 8.6 meV to $\Delta_{1max}$ = 14.5 meV, the peripheral patterns from the inter-pocket scattering have negligible difference while the pattern in the center corresponding to the intra-pocket scattering changes a lot. The magnitude of the outer circle in the central pattern (Figs. 3f and 3g) seems to be enhanced obviously when the energy increases from the small energy gap to the larger one. The intensity of the FT-QPI contour reflects the joint DOS (JDOS) between two *k*-points on the Fermi surfaces. The emergence and enhancement of intensity in the contour in $\vec{q}$ space of the outer circle at a higher energy suggests that the superconducting gap on the poles of the Fermi surface (along the $\Gamma$-$\widetilde{M}$) direction is larger[24-26].

We then pay attention to the central pattern by measuring the QPI mapping in the larger real-space scale to get the detailed feature of the smaller $\vec{q}$-space view, and the evolution of the $Z(\vec{q}, E)$ images is shown in Fig. 4. One can see that there is almost nothing at the locations of the two interference rings at the energy of 0 mV and 2 mV. This may rule out the existence of any gap nodes on the Fermi surface, since the FT-QPI images would show an intensity at the $\vec{q}$-vector connecting two nodal points if they would exist. This is consistent with the full gap structure of the STS spectrum. At an energy of 6.8 mV, a ring like intensity on the FT-QPI images can be observed, with the strongest intensity along the $\Gamma\widetilde{M}$ direction. At the energy of the small gap, namely 8.6 mV, the FT-QPI pattern looks similar to that at 6.8 mV. However, when the energy is further increased to the larger gap, i.e., 14.5 mV, besides the inner ring, a new set of segments appear at larger $\vec{q}$ vectors, which gradually construct an outer ring. When we take the $Z(\vec{q}, E)$ intensity along the contours of the inner and outer rings as shown in Figs. 5a and 5b, an angle dependent $Z(\theta, E)$ at various energies can be obtained. The 2D colour plots of $Z(\theta, E)$ along the inner and the outer rings both show the fourfold symmetry. The low or null intensity of $Z(\theta, E)$ along each ring may suggest that the Fermi surface at this part is still gapped, so both minima of the gap appears in the $\Gamma\widetilde{M}$ directions. Considering the two anisotropic *s*-wave gap functions used in the fitting to the measured STS spectrum, we put the angle dependent gap functions onto the 2D colour plots of $Z(\theta, E)$ as the solid lines in Figs. 5c and 5d. Both gap minima on the two rings appear in the $\Gamma\widetilde{M}$ directions, which is consistent with the direction of gap minima of electron pockets in $FeSe_{0.45}Te_{0.55}$ [27]. In addition, the smaller gap ($\Delta_2$) appears in the inner ring in $\vec{q}$-space.

Concerning the two gap functions with different maximum magnitudes derived from fitting to both STS spectra and the FT-QPI images, we would like to put forward the picture concerning the band hybridization of the two electron pockets in the folded BZ. In iron based superconductors, the electron pockets are usually elliptical-shaped and then reconstruct by hybridization in the folded BZ with two Fe atoms in one unit cell. On the single elliptic Fermi surface, the segments in the pole region are contributed by the $d_{xy}$ orbitals, while the waist areas are derived from the $d_{xz}/d_{yz}$ orbitals[28]. Due to the hybridization, we have two sets of Fermi pockets which correspond well to the inner ($d_{xz}/d_{yz}$) and outer ($d_{xy}$) pockets. The hybridization area consists electrons with mixed orbitals of $d_{xy}$ and $d_{xz}/d_{yz}$. This may be the reason why JDOS is very weak between the overlapped Fermi surfaces and the outer ring seems not to close completely in $\Gamma\widetilde{X}$ directions. If the nodeless d-wave is applicable in the system, sign reversal of superconducting gaps should exist between two neighboring elliptic Fermi pockets. In the hybridized region, the gap should be zero, which is not consistent with our results. In this hybridization picture, our data indicate a smaller gap in the hybridized inner pocket with the $d_{xz}/d_{yz}$ characteristic, while the larger gap accommodates in the outer pocket with the $d_{xy}$ characteristic. To our knowledge, this is the first time to resolve the two sets of Fermi pockets near the $\widetilde{M}$ point. In the situation of *s*-wave gap on the hybridized electron pockets, the theoretically predicted gap maximum locates in the $\Gamma\widetilde{X}$ directions and gap minimum locates in the $\Gamma\widetilde{M}$ directions[13], which is consistent with our results. In our experiment, the spectrum shows full-gaped feature near zero-bias and the local gap minimum locates in the $\Gamma\widetilde{M}$ directions, so the gap distribution may be very similar as the situation shown in Figs. 5c and 5d. Further ARPES experiments with refined resolution are strongly desired to confirm the conclusions derived here. Our observation of two superconducting gaps and the assignment of the two gaps to the hybridized two sets of Fermi pockets in the new superconductor (Li$_{1-x}$Fe$_x$OH)FeSe are very intriguing. Furthermore, the huge gap ratio $2\Delta_1/k_BT_c$ = 8.7 indicates the strong coupling mechanism for superconductivity in the iron based superconductors.

**Acknowledgements**

We acknowledge the useful discussions with Qianghua Wang, Jiangping Hu and Dunghai Lee. This work was supported by the Ministry of Science and Technology of

<p>China (973 projects: 2011CBA00100, 2012CB821403), National Natural Science Foundation of China, and PAPD.</p>

**Author contributions**

The low-temperature STS measurements were finished by ZYD, XY, DLF, GD, HY, and HHW. The samples were prepared by HL and XYZ. The data analysis using the Dynes model were done by ZYD, XY and HY. HHW coordinated the whole work, HY and HHW wrote the manuscript which was supplemented by others. All authors have discussed the results and the interpretation.

**Competing financial interests**

The authors declare that they have no competing financial interests.

*Correspondence and requests for materials should be addressed to

huanyang@nju.edu.cn or hhwen@nju.edu.cn.

Figures and legends

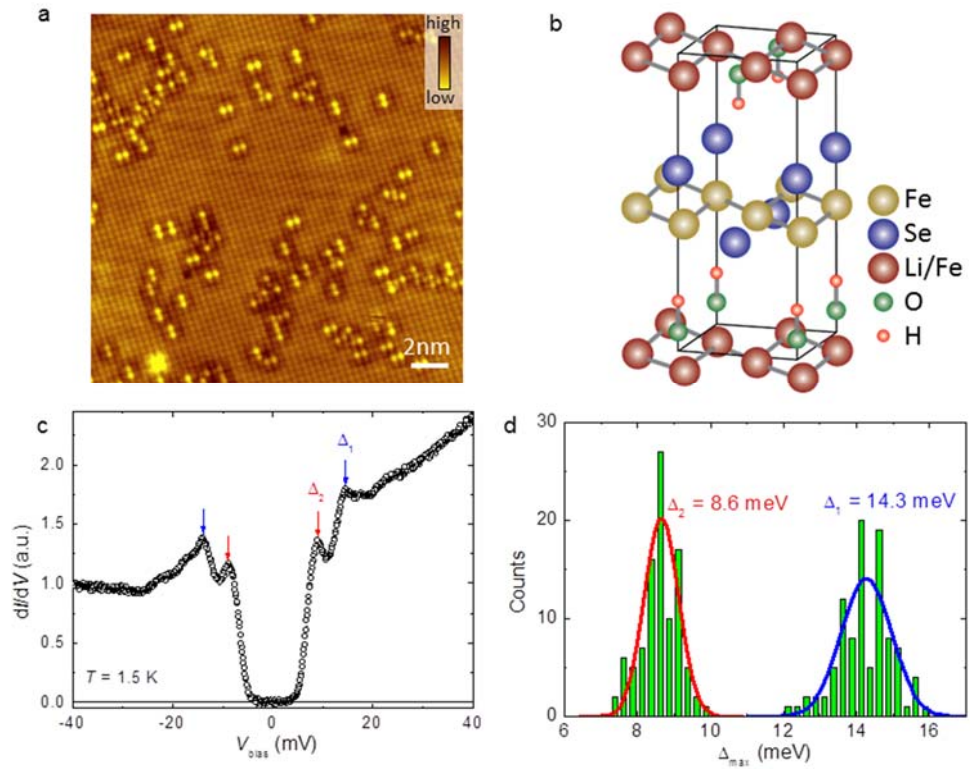

**Figure 1 | Topographic STM image and two-gap feature of (Li$_{1-x}$Fe$_x$OH)FeSe. a**, The atomically resolved STM image with a bias voltage of $V_{bias}$ = 180 mV and tunnelling current of $I_t$ = 102 pA respectively. The defects with the dumbbell shape are observed on the top surface. **b**, The schematic atomic structure of (Li$_{1-x}$Fe$_x$OH)FeSe. **c**, A typical STS spectrum measured at 1.5 K away from the defects. One can clearly see the two-gap feature. **d**, Histogram of the superconducting gap values with the fitting results by Gaussian functions (solid lines).

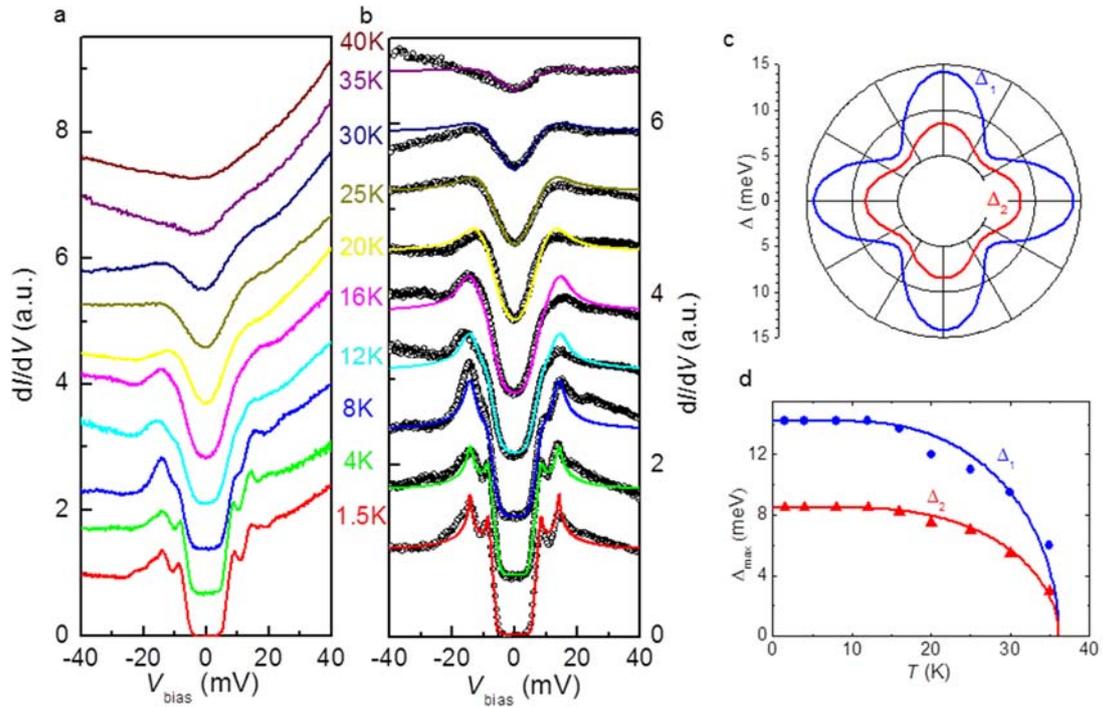

**Figure 2 | Temperature dependent tunnelling spectra and theoretical fits. a**, The evolution of the STS spectra measured at temperatures from 1.5 K to 40 K. **b**, Fitting results to the STS spectra normalized by the one measured at 40 K in the normal state. The black hollow circles represent the experimental data, and the coloured solid lines are the theoretical fits to the data with two anisotropic *s*-wave gaps by the Dynes model. **c**, The anisotropic-gap functions used in the fitting to the curve measured at 1.5 K. **d**, Temperature dependence of the two gaps extracted from the Dynes model fitting, and the solid lines denote the theoretical calculations of the superconducting gap from the BCS model.

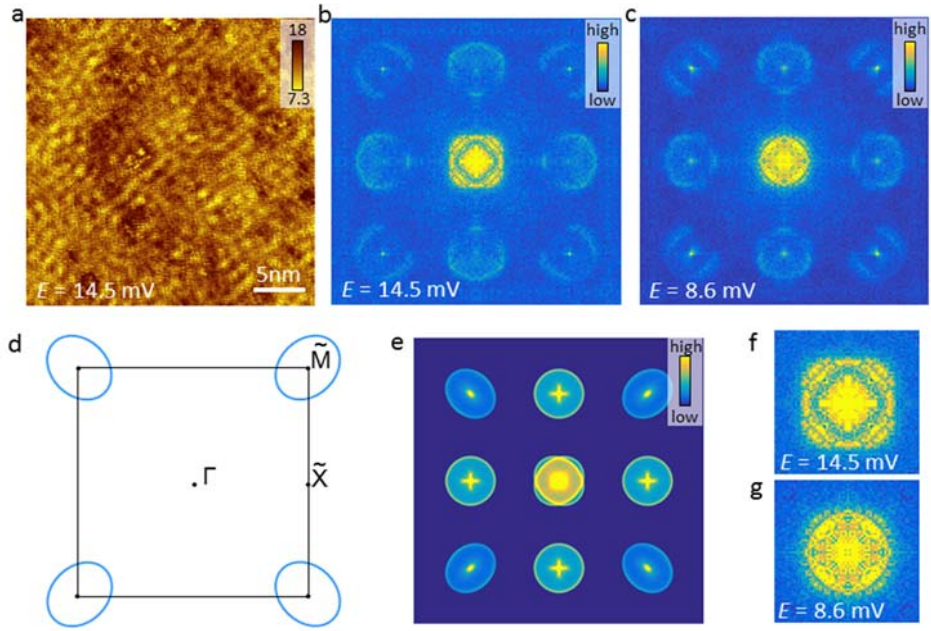

**Figure 3 | Quasiparticle interference patterns and theoretical simulation. a**, QPI image $Z(\vec{r}, E = 14.5$ mV) in real space measured at 1.7 K at the energy of the coherence peak position of the larger gap. **b, c**, $Z(\vec{q}, E)$ obtained by taking Fourier transformations on the corresponding real-space image measured at the coherence peak position of the two gaps, namely 14.5 mV and 8.6 mV. The images are four-fold symmetrized in order to enhance the signal. **d**, Schematic image of the elliptical-shaped electron pockets in unfolded Brillouin zone. **e**, The theoretical simulation of the QPI scattering intensity by applying autocorrelation to **d**. **f, g** Zoom-in images of the central parts of **b** and **c** which contain the information of the small-$q$ intra-pocket scattering. The image at 14.5 mV is very similar to the structure of the central pattern in **e**. However the central pattern at 8.6 mV has very different behavior, and the outer ring seems to be less pronounced.

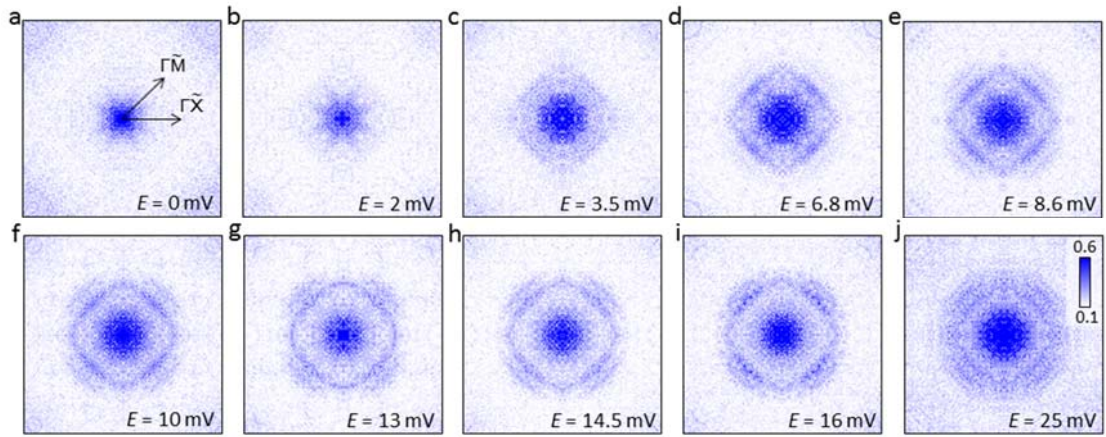

**Figure 4 | The evolution of $Z(\vec{q}, E)$ at different energies.** The FT-QPI $Z(\vec{q}, E)$ images are derived from Fourier transformation on the QPI images with the real space scale of 58 nm × 58 nm. A 2D-gaussian-function background was subtracted from the raw FT-QPI image and the four-fold symmetrization was carried out in order to enhance the signal. The intensity on the inner ring in $\vec{q}$-space shows up at above 3.5 mV, the ring becomes clear at the energy of the smaller gap, namely 8.6 mV; while the segments corresponding to the outer ring appear gradually with a higher energy above 6.8 mV, and are clear when measuring with the voltage at the larger energy gap 14.5 mV.

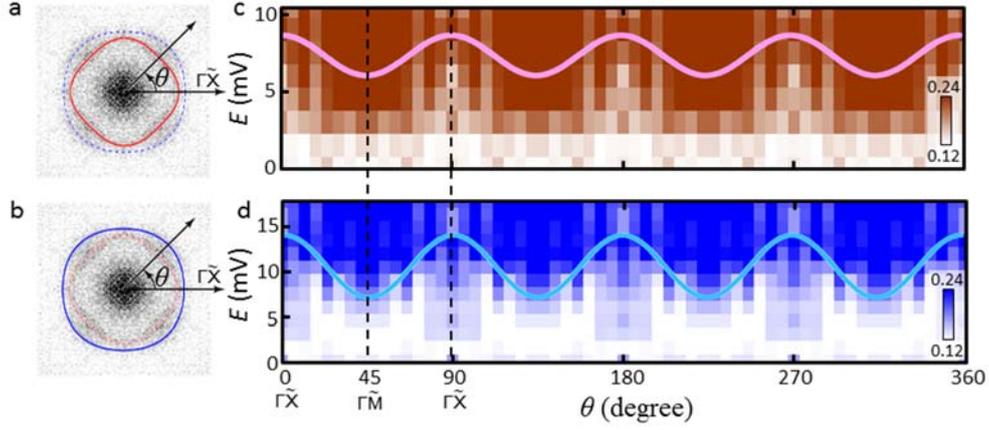

**Figure 5 | 2D colour plots of $Z(\theta, E)$ along the inner and outer contours of the central pattern in $q$-space. c**, **d** The QPI intensity along the inner (**a**) and outer (**b**) Fermi surfaces at different angles taken from the four-fold symmetrized QPI patterns in Fig. 4 at various energy values. The initial polar angle is from the $\Gamma\widetilde{X}$ direction. The gap minimum locates in the $\Gamma\widetilde{M}$ direction for both contours. The solid-line contours in **a** and **b** were expressed approximatively by $R_1-r_1\cos4\theta$ and $R_2+r_2\cos4\theta$ for outer and inner rings respectively. The gap minimum locates in the $\Gamma\widetilde{M}$ direction for both contours. The solid lines in the 2D colour plots are the fitting functions used for the spectrum taken at 1.7 K as shown in Fig. 2**c**.

# SUPPLEMENTARY INFORMATION

**I. Growth and characterization of the single crystal**

The $(Li_{1-x}Fe_xOH)FeSe$ single crystals were synthesized by the hydrothermal ion-exchange method[S1]. Firstly, 6g LiOH (J&K, 99+% purity) was dissolved in 15 mL deionized water in a teflon-linked stainless-steel autoclave (volume 50 mL). Then, 0.6 g iron powder (Aladdin Industrial, 99+% purity), 0.3 g selenourea (Alfa Aesar, 99% purity), and several pieces of $K_{0.8}Fe_{2-x}Se_2$ single crystals (20 to 40 mg in total) were added to the solution. After that, the autoclave was sealed and heated up to 120 °C followed by staying for 40 to 50 hours. Finally, the $(Li_{1-x}Fe_xOH)FeSe$ single crystals can be obtained by leaching.

Figures S1a and S1b show the temperature dependent resistivity and DC magnetization of the $(Li_{1-x}Fe_xOH)FeSe$ single crystals. The sample shows perfect Meissner shielding effect with the superconducting transition temperature around 38 K which is comparable to previous reports. The magnetization-hysteresis-loops shown in Fig. S1c suggest the bulk pinning of vortices and thus the bulk superconductivity of the material. The calculated critical current density at 2 K and 0 T from the Bean critical state model[S2] is one order of magnitude larger than that in $K_xFe_{2-y}Se_2$ and is comparable to that in the optimally doped $Ba(Fe_{1-x}Co_x)_2As_2$[S3].

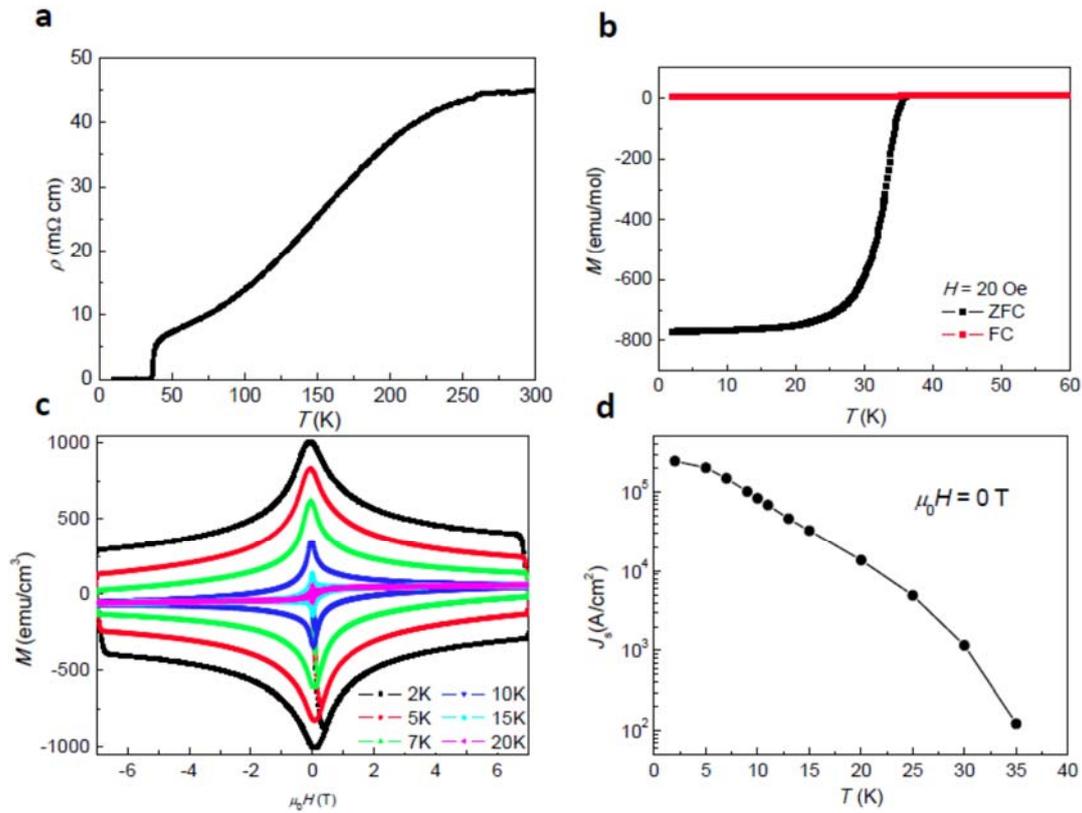

**Figure S1 | Superconducting transition and magnetization of the (Li$_{1-x}$Fe$_x$OH)FeSe crystal. a**, Temperature dependence of resistivity of the sample. **b**, Temperature dependence of DC magnetization measured with the zero-field-cooled (ZFC) and field-cooled (FC) processes at a field of 20 Oe. **c**, The magnetization-hysteresis-loops measured at temperatures from 2 to 20 K. **d**, Temperature dependence of the calculated critical current density $J_s$ from the Bean critical state model at zero magnetic field.

**II. STS spectrum measurements and fitting.**

The STS spectra were measured with an ultrahigh vacuum, low temperature and high magnetic field scanning tunnelling microscope (USM-1300, Unisoku Co., Ltd.). The samples were cleaved in an ultra-high vacuum with a base pressure about 1×10$^{-10}$ torr. During all STM/STS measurements, tungsten tips were used. To lower down the noise of the differential conductance spectra, a lock-in technique with an ac modulation of 0.8 mV at 987.5 Hz was used.

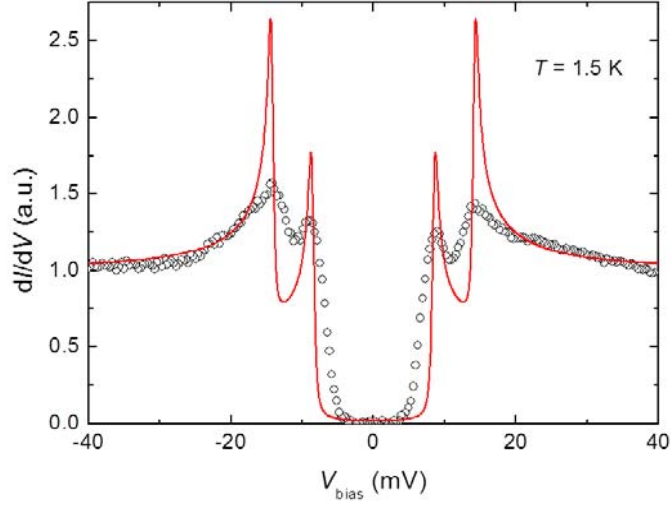

**Figure S2 | Theoretical fit by two isotropic *s*-wave gaps.** The fitting fails to catch the main features of the experimental data.

The tunnelling spectra exhibit a two-gap feature and zero conductance near the zero-bias. So we firstly use two *s*-wave gap to fit the experimental data. In the model to fit the normalized spectra, the tunnelling current is constructed as: $G = x dI_1/dV + (1-x) dI_2/dV$, where $I_{1(2)}(V)$ is the tunnelling current contributed by the larger(smaller) gap, and $p$ is the related spectral weight. $I_{1(2)}(V)$ is given by

$$I_{1(2)}(V) = \int_{-\infty}^{+\infty} d\varepsilon [f(\varepsilon) - f(\varepsilon + eV)] \cdot \mathrm{Re}\left\{ \frac{\varepsilon + eV + i\Gamma_{1(2)}}{\sqrt{(\varepsilon + eV + i\Gamma_{1(2)})^2 - \Delta_{1(2)}^2}} \right\}. \quad (S1)$$

Here, $f(\varepsilon)$ is the Fermi function containing the information of temperature, and $\Gamma_{1(2)}$ is the scattering factor of the larger (smaller) gap, and the result is shown in Fig. S2. One can see that the fitting result fails to catch up the main features of the experimental data, so we instead use two anisotropic *s*-wave gaps to fit the data, and in this situation $I_{1(2)}(V)$ reads

$$I_{1(2)}(V) = \frac{1}{2\pi} \int_{-\infty}^{+\infty} d\varepsilon \int_0^{2\pi} d\theta [f(\varepsilon) - f(\varepsilon + eV)] \cdot \mathrm{Re}\left( \frac{\varepsilon + eV + i\Gamma_{1(2)}}{\sqrt{(\varepsilon + eV + i\Gamma_{1(2)})^2 - \Delta_{1(2)}^2(\theta)}} \right). \quad (S2)$$

We take the anisotropic nodeless gap function $\Delta_{1(2)}(\theta) = \Delta_{1(2)\mathrm{max}}[1 - p_{1(2)}(1 - \cos 4\theta)]$ in which $p$ determines the anisotropy of gaps. In all process of fitting, $\Delta_1(\theta) =$ $\Delta_{1max}(0.25\cos4\theta+0.75)$ and $\Delta_2(\theta) = \Delta_{2max}(0.15\cos4\theta+0.85)$, respectively. The fitting parameters for the normalized spectrum shown in Fig. 2b are listed in Table S1. The spectral weight $x$ of the smaller gap changes at different temperatures.

**Table S1 | Fitting parameters to the experimental data at different temperatures by using two anisotropic *s*-wave gaps.**

| $T$ (K) | $\Delta_{1max}$ (meV) | $\Delta_{2max}$ (meV) | $\Gamma_1$ (meV) | $\Gamma_2$ (meV) | $x$ (weight of $\Delta_1$) |
|---|---|---|---|---|---|
| 1.5 | 14.2 | 8.5 | 0.15 | 0.2 | 55% |
| 4 | 14.2 | 8.5 | 0.15 | 0.15 | 50% |
| 8 | 14.2 | 8.5 | 0.05 | 0.05 | 75% |
| 12 | 14.2 | 8.5 | 0.4 | 0.2 | 75% |
| 16 | 13.7 | 8.2 | 0.3 | 0.2 | 75% |
| 20 | 12 | 7.5 | 1 | 1 | 75% |
| 25 | 11 | 7 | 1.8 | 1.8 | 75% |
| 30 | 9.5 | 5.5 | 2.5 | 2.5 | 75% |
| 35 | 6 | 3 | 3 | 3 | 75% |

**III. Impurity-induced in-gap state.**

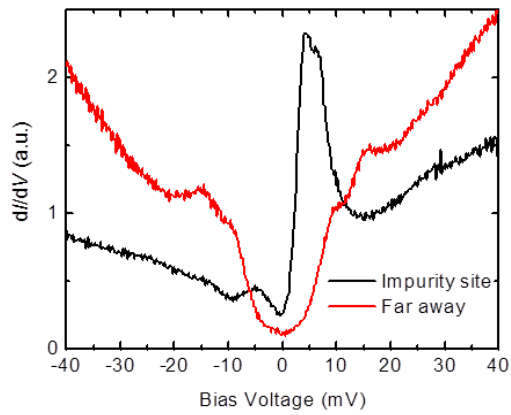

**Figure S3 | STS spectra measured on the impurity site and the clean region far away from impurities.** A typical STS spectrum shown as a red line measured on the center of a dumbbell shape defect shows strong suppression of the coherence peaks as well as clear in-gap impurity states as two resonance peaks with the peak positions at

around ± 5 mV.

**IV. QPI measurements and data treatments.**

We performed the QPI measurements on two areas with different scanning range as their topographies shown in Figs. S4a and S4b. The corresponding QPI images measured at various energies are shown in Fig. 3 and Fig. 4. The large scanning area helps us to investigate the intra-pocket scattering.

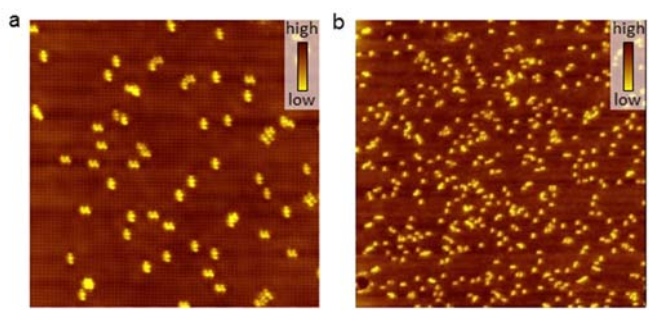

**Figure S4 | Atomically resolved topographies with different scanning range. a**, A topography image of 28 nm × 28 nm (tunnelling condition $V_{bias}$ = 40 mV, $I_t$ = 100 pA). The QPI image on the same area is shown in Fig. 3. **a**, A 58 nm × 58 nm topography image ($V_{bias}$ = 40 mV, $I_t$ = 102 pA) measured on another place, and the QPI image on the same area is shown in Fig. 4.

To reduce the noise of the scattering patterns from the Fourier transformation, we suppress the intensity in the center by multiplying a 2D factor of 1–$A \times g[\vec{q}(0,0),\sigma]$ to the raw FT-QPI images. Here $g[\vec{q}(0,0),\sigma]$ is a 2D Gaussian function with the center at central point $\vec{q}(0,0)$ and the standard deviation of $\sigma$. The parameters $A$ and $\sigma$ were adjusted slightly to make the intensities around a circle far away from the central pattern at different scanning bias voltage the same magnitude. Then we do the fourfold symmetrization to the resultant images. Figure S5 shows the evolution QPI images and the corresponding FT-QPI images at different energy values.

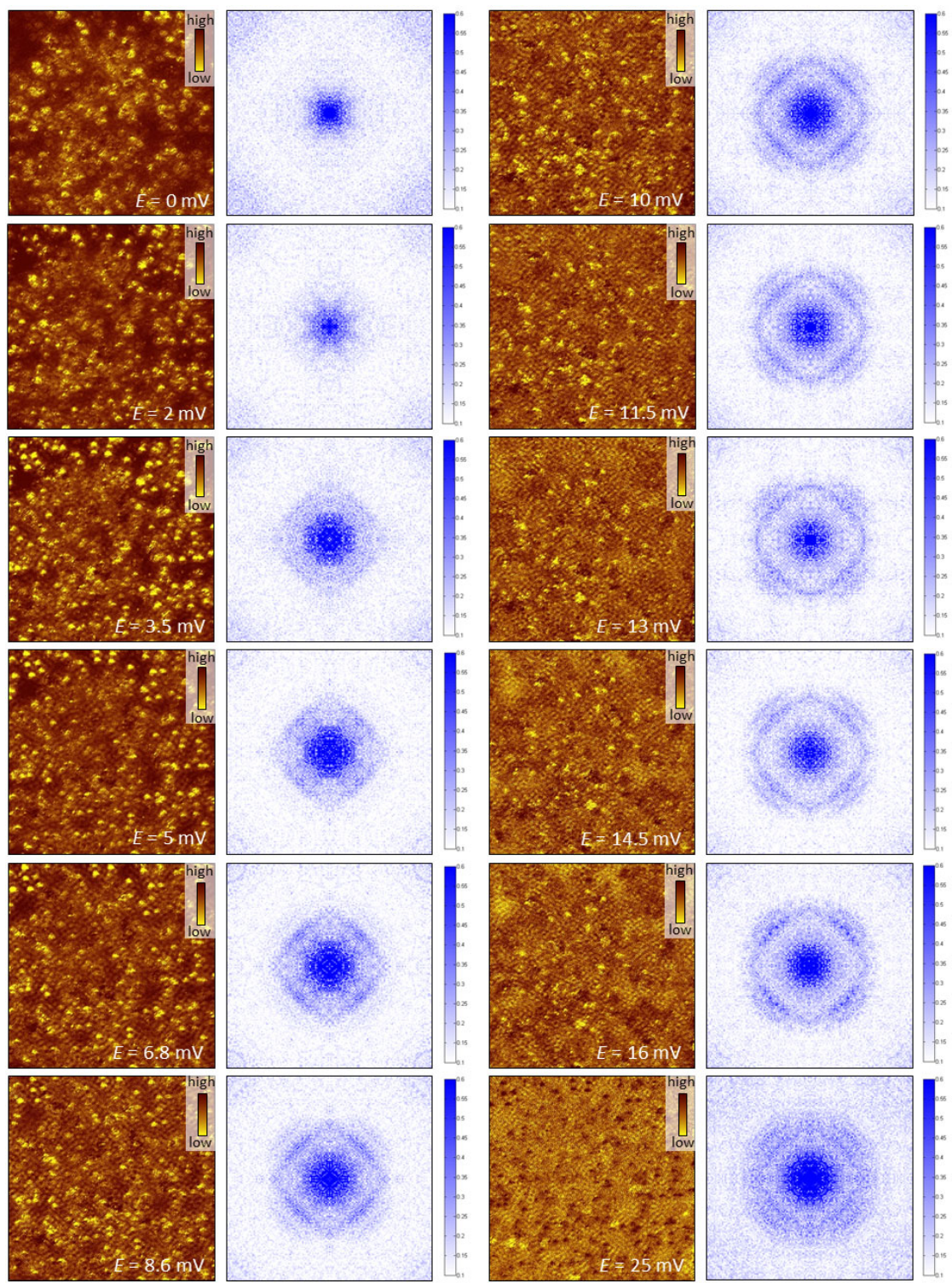

**Figure S5 | Sequence of the QPI images and their respective FT-images.** The QPI images in real space were taken in a 58 nm × 58 nm area.

**Reference:**

S1. Dong, X. L. *et al.* (Li$_{0.84}$Fe$_{0.16}$)OHFe$_{0.98}$Se superconductor: Ion-exchange synthesis of